\documentclass[12pt]{iopart}

\usepackage{setstack}
\usepackage{graphicx}
\usepackage{amssymb}
\begin{document}

\title{Voter Model with Time dependent Flip-rates}

\author{G. J. Baxter}

\address{Departamento de F\'isica, I3N, Universidade de Aveiro,
Campus Universit\'ario de Santiago, 3810-193 Aveiro, Portugal}
\ead{gjbaxter@ua.pt}

\begin{abstract}
We introduce time variation in the flip-rates of the Voter Model. 
This type of generalisation is relevant to models of ageing in language
change, 
allowing the representation of changes in speakers' learning rates
over their lifetime.
and may be applied to any other similar model in which
interaction rates at the microscopic level change with time. 
The mean time taken to reach consensus varies in a nontrivial way with
the rate of change of the flip-rates, 
varying between bounds given by the mean consensus times for
for static homogeneous (the original Voter Model) and static
heterogeneous flip-rates.
By considering the mean time between interactions for each agent, we
derive excellent estimates of the mean consensus times 
and exit probabilities 
for any time scale of flip-rate variation.
The scaling of consensus times with population size on complex
networks is correctly predicted, and is as would be expected for the
ordinary voter model.
Heterogeneity
in the initial distribution of opinions has a strong effect,
considerably reducing the mean time to consensus, while increasing the
probability of survival of the opinion which initially occupies the
most slowly changing agents.
The mean times to reach consensus for different states are very
different. An opinion originally held by the fastest changing agents
has a smaller chance to succeed, and takes much longer to do so than
an evenly distributed opinion.
\end{abstract}

\vspace{2pc}
\noindent{\it Keywords}: Stochastic processes, Population dynamics
(Theory), Interacting agent models
\maketitle


\section{Introduction\label{introduction}}

Neutral diffusion-like or copying process models have been applied in a
very broad range of fields, from social phenomena
\cite{Castellano2009}  and language change \cite{Baxter06,Baxter09}, to ecology
\cite{Horvat2010,Hubbell2001,Condit02} and population genetics
\cite{Crow70} among many
more. In all such models, different alternative items -- species,
opinions or language variants for example -- are copied between
neighbouring sites
until one finally dominates the whole system.
Here we consider the effect of time variation in the rate of update
at each site in such models. 
This has particular relevance to language change, in
which the rate at which speakers adapt to their language environment
changes with age. The items copied are alternative variants of a
language element; different ways of `saying the same thing'. Young
speakers adapt very quickly, but once they
reach adulthood many speakers barely change their language use
\cite{Bailey1991,Sankoff2006}.
We show that the
time to reach consensus depends in a non trivial way on the time scale
of the variations of agent update rate, see fig.~\ref{fixtime_vs_S}.
By considering the mean time between interactions for each agent, we
are able to obtain excellent estimates of the mean time to reach
consensus, even in the
intermediate regime where the timescale of consensus formation and the
time scale of flip-rate change are of the same order, and interact in
a non-trivial way.

The generalisation and the methods we will describe could be
applied to any of the models mentioned, but for simplicity we
concentrate on the Voter Model \cite{Liggett99}, in which agents in a
population possess one of two discrete opinions.
At each step an agent is chosen and imports the opinion of a randomly
selected neighbour. The rate at which a particular agent is chosen for
update is their flip-rate, and it is time variance of these flip-rates
that we consider in this paper.
The Voter Model has become an emblematic opinion
spreading model due
to its simplicity and tractability, as well as its distinction from other
coarsening phenomena such as the Ising model \cite{Dornic2001}.
The original Voter Model has been extended to include network
structure \cite{Sood05,Suchecki05b, Baronchelli2010} and the
effect of changes in the microscopic interactions \cite{Sood08,
  Schneider-Mizell2009}. 
Masuda {\it et al.} investigated the effect of heterogeneity in the
flip-rates of agents \cite{Masuda2010}. 
While they have some spirit in common, the present model differs from other 
time dependent models including such effects as latency or ageing of states
\cite{Stark2008,Lambiotte2009, FernandezGracia2011} in that the change
in flip rate doesn't
depend on the opinion or the time of adoption of the opinion. That is,
we are not interested in ageing of the opinions but of the agents
themselves. The present work is probably most similar to the
`exogenous update' rule of \cite{FernandezGracia2011}, however in that
case the update rule is the same for all agents and is reset when an
agent becomes available for update, rather than changing independently
.

In the heterogeneous Voter Model \cite{Masuda2010}, a population of
$N$ agents (labelled $i = 1,2,...,N$) possess opinions $x_i$ which
take values either $1$ or $0$. Agents are selected for update asynchronously
with frequency proportional to flip-rates $r_i$, which may be
different for each agent. Update consists of an agent importing the
opinion of a neighbour, selected uniformly at random. 
In the present study, we extend this model to consider the flip-rates
to be time dependent, thus $r_i(t)$.
The method
described here is perfectly amenable to considering complex network
structure. For simplicity, we first
consider a well mixed population (that is, a fully connected network)
before examining more complex structures.

The mean time to consensus for homogeneous flip-rates $r_i=r,\; \forall
i$, that is, the standard Voter Model, is well established
\cite{Sood05,Baxter08,Blythe2010}, and is found by assuming the population of
agents relaxes quickly to a quasi stationary state (QSS), followed by
a much longer period characterised by collective motion. The mean time
to consensus can be calculated for this second stage by writing a
Fokker-Planck equation for a conserved centre-of-mass variable. This gives a
lower bound and a good approximation to the total mean time to consensus. When
heterogeneity is introduced in the flip-rates, the consensus time is
always increased, and can be predicted from the moments of the
flip-rate distribution \cite{Masuda2010}.
For
very slowly changing flip-rates, the consensus time is essentially the
same as for the static heterogeneous case. For very quickly changing
flip-rates, the consensus time may be reduced to the value found for
homogeneous flip-rates with the same mean. For intermediate periods,
the time to consensus varies smoothly between these two extremes, as
can be seen in fig.~\ref{fixtime_vs_S}.
Using the
distribution of flip-rates existing in the population simply returns
the static heterogeneous result, which doesn't vary with the period of
variation of the
flip-rates. Instead, we calculate effective flip-rates,
found by considering the mean time between interactions for each
agent. Using the moments of the distribution of these effective
flip-rates returns the correct qualitative behaviour, and is in
excellent quantitative agreement with numerical results
(fig.~\ref{fixtime_vs_S}).

\section{Analysis\label{analysis}}

The mean time to consensus can be calculated through a Fokker-Planck
Equation (FPE) formalism. See for example \cite{Baxter08,Sood05} or
the rigorous treatment in \cite{Blythe2010}.
The essential idea is that after an initial, rapid, period of mixing,
the individual opinions settle into a long lived meta-stable
distribution. This quasi-stationary state (QSS) is characterised by a
weighted mean opinion which is conserved by the dynamics. The
mean time to consensus can be calculated by considering the evolution
of only this central variable.

\subsection{Mean consensus time for static flip-rates}

 For orientation, we first
calculate the mean consensus times for static homogeneous (i.e. the
basic Voter Model) and static heterogeneous flip-rates.
We define a weighted mean opinion $\xi(t)$ by:
\begin{equation}\label{xi_static}
\xi(t) \equiv \frac{\sum_i {x_i(t)}/{r_i}}{\sum_i {1}/{r_i}}
 = \sum_iQ_ix_i(t)\;
\end{equation}
where
\begin{equation}\label{QSS_static}
Q_i \equiv \frac{1}{N\overline{({1}/{r})}}\; \frac{1}{r_i}
\end{equation}
and $\overline{(\;\cdots\;)}$ signifies a population average.
We choose $\xi$ because it is conserved by the
dynamics \cite{Castellano2005, Suchecki05, Sood08, Baxter08}:
\begin{equation}
\frac{{\rm d}}{{\rm d}t}\langle \xi(t)\rangle = 0.
\end{equation}
This also means that the probability that the population eventually
reaches consensus in the state ${\bf 1}$ is simply
\begin{equation}
P({\bf x} \to {\bf 1}) = \xi(0) \equiv \xi_0.
\end{equation}

The fraction $x(t) \equiv \frac{1}{N}\sum_ix_i(t)$
of agents in the population holding opinion $1$, equivalently the
`magnetisation' converges rapidly to $\xi_0$:
\begin{equation}
\frac{{\rm d}\langle x\rangle}{{\rm d} t} = \langle x\rangle - \xi_0\,,
\end{equation}
as do in turn the expected values of the individual opinions $x_i$:
\begin{equation}
\frac{{\rm d}\langle x_i\rangle}{{\rm d} t} =
\frac{r_i(t)}{Nr_0}[\langle x_i\rangle - \langle x\rangle]\,.
\end{equation}
After this initial mixing the system is in
a long-lived quasi-stationary state (QSS) in which the individual
opinions are subordinated to the centre-of-mass variable $\xi(t)$,
which changes only very slowly. 
The QSS is found by setting $x=\xi$ and calculating the distribution
of the $x_i$ about a fixed $\xi$. 
We can calculate the mean time, $T^*$,  to reach consensus beginning
from this QSS by considering only the central variable $\xi(t)$.

The conservation of $\langle\xi\rangle$ means that the FPE for the
probability distribution of $\xi(t)$ has only a diffusion term,
originating from the second jump moment:
\begin{equation}
\langle\delta\xi^2\rangle =
\frac{1}{\bar{r}\overline{(\frac{1}{r})}N(N-1)}\left[\xi(1- x) +
  x(1-\xi)\right]\,.
\end{equation}
Choosing the time
increment $\delta t = 1/N$, in the limit of large $N$ we arrive at
\begin{equation}\label{FPE_xi_static}
\frac{\partial}{\partial t}P(\xi,t') =
\frac{1}{\bar{r}\overline{(\frac{1}{r})}N}
\frac{\partial^2}{\partial \xi^2}\left[\xi(1-\xi)P(\xi,t')\right]\;,
\end{equation}
where we 
have used the fact that in the QSS $x = \xi$.
We use $t'$ to emphasise that the QSS is now taken as the
starting point. Thus $t'=0$ when we first arrive in the QSS (at some
time $t^*$, that is $t' = t-t^*$).

Note that the state variables $x_i$ are discrete. It is however also
possible to write a complete FPE in continuous variables, either by
assuming a
large population and aggregating all agents with $r_i$ in the range
$[r_i,r_i+\delta r)$ \cite{Sood05}, or by considering each agent to be
  occupied by a number $M$ of particles. For $M\to\infty$, $x_i$
  becomes a continuous variable. It was shown in \cite{Blythe2010}
  that if the rate of exchange of particles between agents is much
  slower than the copying within an agent, the results for large $M$
  apply to the original case $M=1$.

The mean time taken to reach consensus then obeys the backward-FPE
\cite{Gardiner}
\begin{equation}\label{T_BFPE_static}
-1 = \frac{1}{\bar{r}\overline{(\frac{1}{r})}N}
\xi_0(1-\xi_0)\frac{{\rm d}^2}{{\rm d}\xi_0} T^*\;.
\end{equation}
Because $\xi$ is conserved, we use the initial value $\xi_0$ which is
equal to the expected value of $\xi(t^*)$ in the QSS. In the fully
connected, and indeed in many other networks
we might consider \cite{Baxter08,Blythe2010}, $t^* \ll T$, so we can
use $T^*$
as a proxy for the overall mean time to consensus $T$.
The mean flip-rate $r_0$ merely determines an overall time scale,
so we can, without loss of generality, also set $r_0=1$.
Furthermore, for large populations we can use the moments $\mu_k$ of
$p(r)$ in place of population averages $\overline{r^k}$, so that
$\overline{(1/r)} \to \mu_{-1}$.
Solving eq. (\ref{T_BFPE_static}) then gives
\begin{equation}\label{T_hetero_static}
T_{\mbox{het}} \approx N \mu_{-1} \left[\xi_0 \ln \xi_0 + (1-\xi_0) \ln(1-\xi_0)\right]\;.
\end{equation}
This agrees with the result obtained previously in \cite{Masuda2010}.

In the homogeneous flip-rate case, $\xi_0 = x_0$, and  $\mu_{-1} =
1/\mu_1 = 1$, so that we recover the standard result
\begin{equation}\label{T_homo_static}
T_{\mbox{hom}} \approx N \left[x_0 \ln x_0 + (1-x_0) \ln(1-x_0)\right]\;.
\end{equation}

\subsection{Time-dependent flip-rates\label{time_dependent}}

Now we are ready to consider the case where the flip rates $r_i(t)$
can vary with time.
We assume that the flip-rates $r_i(t)$ follow
some periodic function with the period $S$ acting as a control
variable.
Initial values for $r_i$ are chosen by
  selecting $s_i$ uniformly at random from $[0,1)$
  and setting $r_i(0) = f(s_i)$ for some $f(s)$.
It is convenient to ensure a stable distribution of $r_i$ values in
the population over time. In the language change application, this
corresponds to a stable distribution of speaker ages in a
population, with old speakers periodically replaced by young
ones, and whose members' learning rates all follow the same function
of age [i.e. $r(s)$]. This is obviously a very crude model, and our
aim here is simply
to demonstrate the general effect of time variation in such
interactions. 
To achieve this, we define a periodic
version of $f(s)$: let $r(s) \equiv f(s - \lfloor s\rfloor)$.
For an agent with initial flip rate $r_i(0) = r(s_i)$, we set
\begin{equation}\label{ri}
r_i(t) = r\left(s_i + \frac{t}{S}\right)\;.
\end{equation}
This ensures the period is equal to $S$ and also that at any time, the
overall distribution of $r_i(t)$ values in a large population follows
$p[r(s)]{\rm d}r= {\rm d}f(s)/{\rm d}s$.

\begin{figure}[tb]
\centering\includegraphics[width=0.70\textwidth]{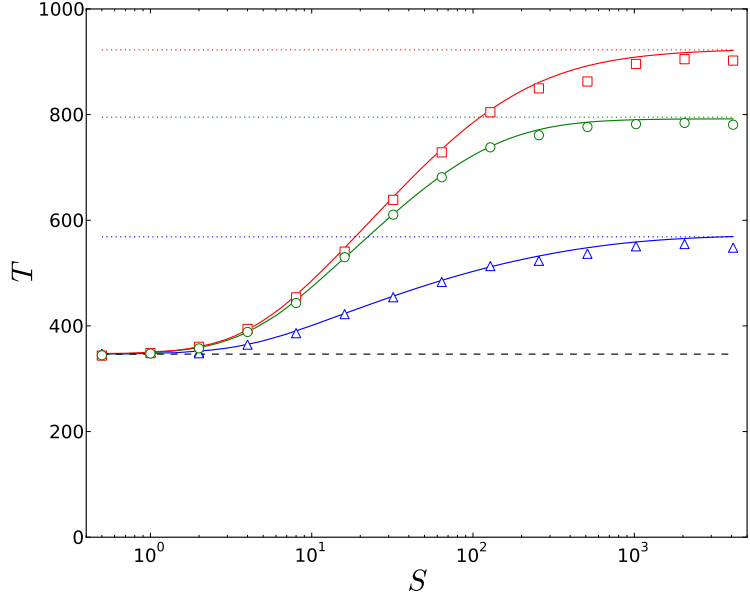}
\caption{Mean consensus time $T$ as a function of period $S$ of change
  of flip-rates for three
  different flip rate functions $r(t/S)$. Numerical simulations are
  shown as open symbols for sinusoidal function $r_{\mbox{sin}}$
  (circles), linear `sawtooth' $r_{\mbox{lin}}$ (triangles) and
  quadratic sawtooth $r_{\mbox{qdr}}$ (squares). Solid curves are
  analytic calculations using eq.~(\ref{T_interp}). 
Dashed line is mean consensus time $T_{\mbox{hom}}$ for homogeneous
flip-rates, dotted lines for heterogeneous but static flip-rates,
$T_{\mbox{het}}$.
Simulation results are averages over $10^4$ runs with homogeneous
initial conditions and population size $N = 500$.
}\label{fixtime_vs_S}
\end{figure}

We postulate that the change in observed consensus time is due to an
interaction between the time scales of flip-rate change and of opinion
change (or consensus formation) of the population. When the flip-rates
change extremely slowly, consensus will be reached with essentially no
change in flip-rates, hence the static heterogeneous result
eq.~(\ref{T_hetero_static})
applies. If the flip-rates change extremely quickly, agent $i$ will
cycle through all the possible values of $r_i$ in a very short time
compared with the rate at which she interacts. We can therefore
calculate an approximate consensus time by replacing $r_i$ with
$\langle r_i\rangle_t=r_0$. We see, then, that
in the limit of very quickly changing flip-rates
$T$ is given by
eq. (\ref{T_homo_static}) i.e. the consensus time in the standard
homogeneous Voter Model.
The heterogeneous flip-rate consensus time (\ref{T_hetero_static}) and the
fast change limit (\ref{T_homo_static}) provide approximate upper and lower
bounds for the consensus time. As can be seen in fig.
\ref{fixtime_vs_S}, these bounds agree very well with numerical results in the
two limits, and consensus times for intermediate regimes lie between
the two.
 
We can calculate a more rigorous interpolation between the results
(\ref{T_hetero_static}) and (\ref{T_homo_static}) by
observing that in both cases, the weight for each agent's
state in the sum for $\xi(t)$ in eq.~(\ref{xi_static}) is proportional
to $1/r_i$, which is the
expected time interval between interactions for agent $i$.
For intermediate $S$, let us generalise by defining $\tau_i(t)$ to be the
expected interval
between interactions for agent $i$, where the dependence on $t$
indicates that this update interval varies because $r_i(t)$ varies
with time.
Consider a sequence of short intervals of length $\Delta t$, beginning
at time $t$. The probability that $i$ is selected in the first
interval is $r_i(t)\Delta t$. The probability that $i$ is selected in
the second (having not been selected in the first) is $[1-r_i(t)\Delta
  t]r_i(t+\Delta t)\Delta t$, and so on. The probability that $i$ is
selected in the $(k+1)$-th such interval is thus:
\begin{equation}\label{litprob_discrete}
r_i(t+k\Delta
   t)\Delta t\prod_{l=0}^{k-1}[1-r_i(t+l\Delta t)\Delta t]\;.
\end{equation}
Taking $\Delta t\to 0$ we can write the terms in the product as exponentials,
i.e. $1-r_i(t+l\Delta t)\Delta t \to \exp\{-r_i(t''){\rm d}t\}$, where
we have rewritten 
$t+k\Delta t \to t'$ and $t+l\Delta t \to t''$. The
product then becomes an integral in the argument of the exponential,
so that the probability that $i$ is selected in the interval
$[t',t'+{\rm d}t)$  becomes
\begin{equation}
r_i(t'){\rm d}t \exp\left\{-\int_t^{t'}r_i(t''){\rm d}t'' \right\}\;.
\end{equation}
Multiplying by the waiting times $(t'-t)$ and integrating gives the
expected waiting time
\begin{equation}\label{tau_integral}
\tau_i(t) = \int_t^{\infty} (t'-t)r_i(t')
\exp\left\{-\int_t^{t'}r_i(t''){\rm d}t'' \right\} {\rm d}t'\,.
\end{equation}
We can then define an effective flip-rate
\begin{equation}\label{r_eff}
\tilde{r}_i(t) \equiv \frac{1}{\tau_i(t)}.
\end{equation}
For the static case, we recover $\tilde{r}_i(t) = r_i$.
For extremely quickly varying $r_i(t)$, 
$\tilde{r}_i(t) = \bar{r}$,
which is the same value as obtained in the original homogeneous Voter
Model. 
To see this, notice that fluctuations in $\int_t^{t'}r(t''){\rm d}t''$ die
out very quickly, so after some small time $\sigma$, 
$\exp\{-\int_t^{t'}r_i(t''){\rm d}t''\} \approx \exp\{-(t'-t){\bar{r}}\}$, while
for times less than $\sigma$, the exponential term is
close to $1$ and $\int_t^{t+\sigma} (t'-t)r_i(t'){\rm d}t' \approx \sigma^2
{\bar{r}}$,
giving $\tau_i(t) \approx \sigma^2 r_o + e^{-\sigma {\bar{r}}}(1+\sigma
{\bar{r}})/{\bar{r}} \approx 1/{\bar{r}}$.
 These limits agree with the two limits obtained through the qualitative
 arguments above.

For the reduction to a single variable, we then define
\begin{equation}\label{xi_tilde}
\tilde{\xi}(t) \equiv \frac{\sum_i \tau_i(t)x_i(t)}{\sum_i \tau_i(t)}\;.
\end{equation}
The argument proceeds as before, so that we in effect
replace $\overline{(1/r)}$ by $\overline{\tau_i(0)}$
and $\bar{r}$ by $\tilde{r}_0\equiv \overline{\tilde{r}_i(0)}$
in eq.~(\ref{T_BFPE_static}), to give
\begin{equation}\label{T_BFPE_full}
-1 = \frac{1}{N \overline{\tau_i(0)}\tilde{r}_0} \tilde{\xi}_0(1-\tilde{\xi}_0)
\frac{{\rm d}^2}{{\rm d}\tilde{\xi}_0}T^*\;.
\end{equation}
For homogeneous initial conditions, $\tilde{\xi}_0 = x_0$.
We can utilise the fact that all agents follow the same flip-rate function
$r(s)$, but starting at different initial values of $s$ to estimate
$\overline{\tau_i(0)}$. In the
large-$N$ limit, the agent's initial values of $s_i$ evenly populate
the interval $[0,1)$. For large $N$, then, we can replace the average
  over agents by an average over $s$, giving
\begin{equation}\label{tau_estimate}
\overline{\tau_i(0)} \approx \int_0^1 \tau(s) {\rm d}s \equiv \tau_0\;,
\end{equation}
where $\tau(s) \equiv \tau_i(0)$ for $i$ such that $r_i(0) = r(s)$.
Finally then we can write 
\begin{equation}\label{T_interp}
T \approx N\tau_0\left[x_0 \ln x_0 + (1-x_0) \ln(1-x_0)\right]
\end{equation}
for homogeneous initial conditions.
This analytic calculation is in excellent agreement with
numerical results for various $r(s)$ distributions over the whole range
of $S$, as can be seen in fig.~\ref{fixtime_vs_S}.

Higher moments can be calculated iteratively using equations \cite{Gardiner}
\begin{equation}\label{higher_moments}
-nT_{n-1} = \frac{1}{N \tau_0
} \tilde{\xi}_0(1-\tilde{\xi}_0)
\frac{{\rm d}^2}{{\rm d}\tilde{\xi}_0}T_n\;,
\end{equation}
where $T_n$ is the $n$\textsuperscript{th} moment of the consensus
time distribution
leading to expressions in terms of polylogarithm functions. 
The
variance of consensus times calculated in this way is in excellent
agreement with numerical simulations (not shown).
In fact eq.~(\ref{higher_moments}) only differs from the ordinary
voter model by a factor $\tau_0$, so
the whole distribution of consensus times has the same shape as that
for the ordinary voter model, once time is rescaled by
$\tau_0$. This is borne out in simulations, see fig.~\ref{T_distn}.

\subsection{Network Structure\label{network}}

In \cite{Sood08} it was shown that a similar mean-field approach is
sufficient to reproduce the population size scaling of the mean
consensus time on heterogeneous networks for the ordinary voter
model. By assuming flip-rates to be independent of degree, a similar
treatment can be performed here. We now show that the population size
scaling for
time varying flip-rates (and hence also for the heterogeneous
flip-rate model of ref. \cite{Masuda2010}) depends on the network
degree distribution in the same way as in the ordinary voter model.

We define the weighted mean to be
\begin{equation}\label{xi_network}
\xi(t) \equiv \frac{\sum_i {x_i(t) q_i}/{r_i}}{\sum_i {q_i}/{r_i}}
\end{equation}
where $q_i$ is the degree of voter $i$. Because the probability that
voter $i$ is chosen for update is proportional to $1/q_i$, we again
find that $\xi$ is conserved by the dynamics.
Carrying out averages over $q_i$ and $r_i$ separately (they are chosen
independently), and, as before, replacing population averages with
distribution moments we find, for large populations,
\begin{equation}
\langle\delta\xi^2\rangle \approx
\frac{1}{r_0\mu_{-1}\langle q\rangle^2N^3}\sum_q n_q q^2\left[x(1- x_q) +
  x_q(1-x)\right]\,.
\end{equation}
Where $n_q$ is the number of voters having degree $q$, and $q$ is the
mean opinion of such voters.
Finally, in the QSS we set $x_q = x = \xi$ giving
\begin{equation}\label{xi2_network}
\langle\delta\xi^2\rangle \approx
\frac{\langle q^2\rangle}{r_0\mu_{-1}\langle q\rangle^2N^2}\xi(1-\xi)\,.
\end{equation}
This differs from the fully connected result only by a factor $\langle
q^2\rangle/\langle q \rangle^2$, and hence the mean consensus time for
heterogeneous flip-rates on a network goes as
\begin{equation}\label{T_static_netw}
T \propto N \mu_{-1}\frac{\langle q\rangle^2}{\langle q^2\rangle}\;.
\end{equation}
It follows that for time varying flip-rates and homogeneous initial
conditions, the mean consensus time goes as [compare Eq. (\ref{T_interp})]:
\begin{equation}\label{T_interp_netw}
T \propto N \tau_0\mu_{-1}\frac{\langle q\rangle^2}{\langle q^2\rangle}\;.
\end{equation}
As can be seen in fig.~\ref{networks_scaling}, this agrees with
numerical simulation for Erdos-Renyi networks and for scalefree
networks.

\subsection{Inhomogeneous Initial Conditions\label{heteroIC}}

We now consider the effect of heterogeneity in the initial opinions of agents.
For the static heterogeneous case, the probability $P({\bf x}\to
{\bf1})$ to reach consensus
state ${\bf 1}$ also depends on the flip-rates of the agents. The opinion
$1$ is more likely to achieve consensus if it initially occupies the
more slowly changing agents, and {\it vice-versa}. 

Returning to eq.~(\ref{xi_tilde}), $\langle\tilde{\xi}\rangle$ is a conserved
quantity to the extent that the replacement of $r_i(t)$ by
$\tilde{r}_i(t)$ in the Fokker-Planck Equation (\ref{FPE_xi_static}) is a
valid approximation. The probability to reach consensus in the state
${\bf 1}$ is then simply $\tilde{\xi}_0$. 
This probability varies with the period of change $S$ of the
flip-rates. When the flip-rates change very fast, the agents are
essentially identical (having effective flip rate $\bar{r}$) and so there
is no initial configuration dependence. Conversely, the effect of
initial inhomogeneity in opinion (that is, correlation between initial
flip-rate and initial opinion) will be strongest for extremely slowly
varying flip-rates. This can be seen in fig.~\ref{fixprob_vs_S}.

The mean consensus time again obeys eq. (\ref{T_BFPE_full}), but the
presence of inhomogeneous initial conditions is felt through the fact that now
$\tilde{\xi}_0 \neq x_0$, leading to
\begin{equation}\label{T_inhomog}
T \approx N\tau_0 \left[\tilde{\xi}_0 \ln \tilde{\xi}_0 +
  (1-\tilde{\xi}_0) \ln(1-\tilde{\xi}_0)\right]\;.
\end{equation}
This means we must calculate $\tilde{\xi}_0$ using the
initial distribution of $x_i$. In the examples shown here this was
done by integrating eq.~(\ref{tau_integral}) numerically and then
computing eq.~(\ref{xi_tilde}) as an integral over $s$ [compare
  eq. (\ref{tau_estimate})] to find $\tilde{\xi}_0$.
The value of $\tau_0$ depends only in $r(s)$ and $S$, so is
independent of initial conditions.
As for the homogeneous case, $\tau_0$ increases with $S$,
and because the fastest change in $\tau_0(S)$ and $\tilde{\xi}_0(S)$
occur in different ranges of $S$ (compare fig.~\ref{fixtime_vs_S} with
fig.~\ref{fixprob_vs_S})
the final curve for $T(S)$ has the complicated
shape seen in fig.~\ref{fixtime_vs_S_inhomog}, borne out in 
simulation.

To calculate the mean time to consensus at a particular state, we
again solve eq.~(\ref{T_BFPE_full}) but the appropriate boundary
conditions are different. This gives
\begin{eqnarray}
T_0 &\approx N\tau_0 \frac{\tilde{\xi}_0}{(1-\tilde{\xi}_0)} \ln
\tilde{\xi}_0\;,\label{T0}\\
T_1 &\approx N\tau_0 \frac{(1-\tilde{\xi}_0)}{\tilde{\xi}_0}
\ln(1-\tilde{\xi}_0)\;.\label{T1}
\end{eqnarray}
Combining these two times in proportion to the probability to reach
each state returns eq.~(\ref{T_inhomog}).

\section{Numerical Simulations\label{numerical}}


We performed simulations of the model
with different distributions of $r_i$. 
We considered the following functional forms for $r(s)$, chosen to
give some variety of interesting functional forms:
\begin{eqnarray}
r_{\mbox{qdr}}(s) &= r_0-r_d + 3r_d(1-s+\lfloor s\rfloor)^2,
&\qquad\vcenter{\hbox{\includegraphics[width=0.09\textwidth]{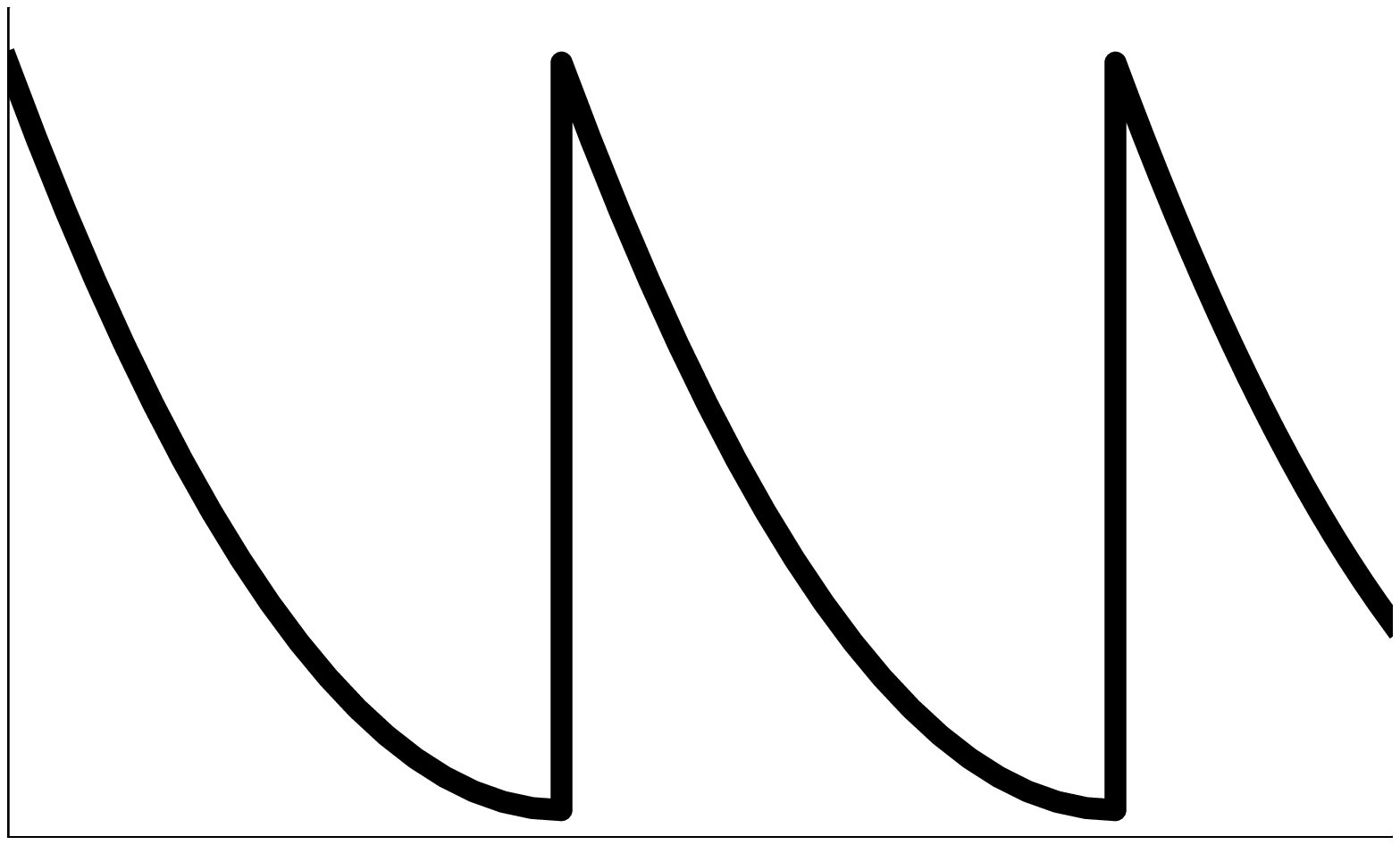}}}\\
r_{\mbox{sin}}(s) &= r_0 + r_d\sin(2\pi s),
&\qquad\vcenter{\hbox{\includegraphics[width=0.09\textwidth]{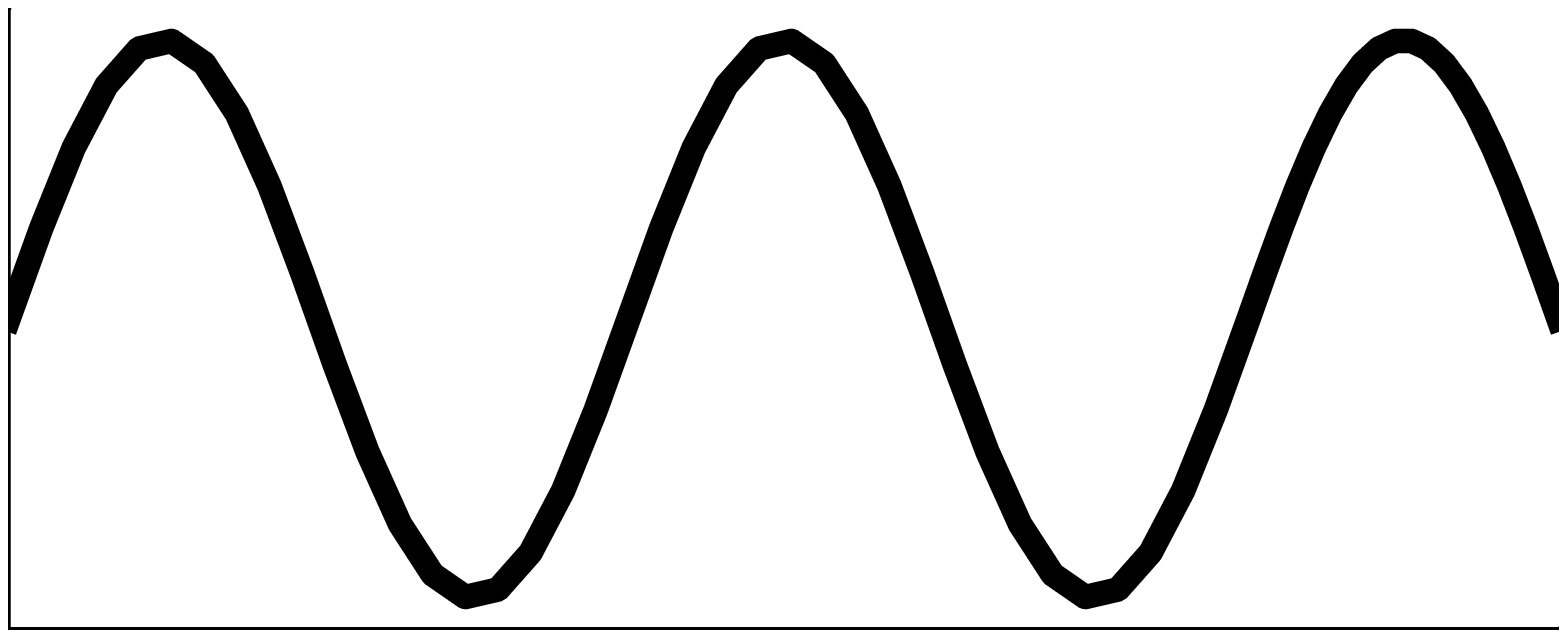}}}\\
r_{\mbox{lin}}(s) &= r_0-r_d + 2r_d(s-\lfloor s\rfloor).
&\qquad\vcenter{\hbox{\includegraphics[width=0.09\textwidth]{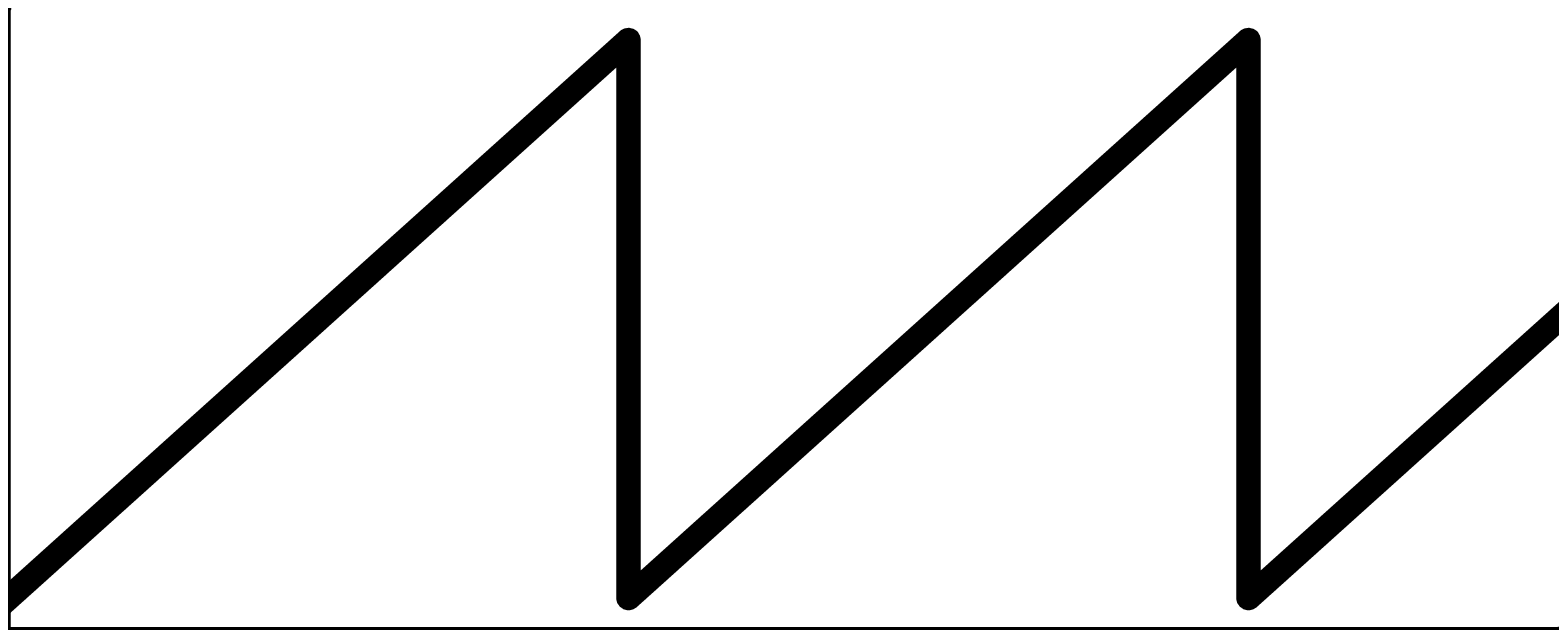}}}
\end{eqnarray}
The mean of each function is $\mu_1 = r_0$, which we generally choose
to be equal to $1$. The parameter $r_d$
controls the amplitude of the variations. To ensure the values of
$r_i$ are always strictly positive, we require $r_d < r_0$.
The mean time to consensus and probability to reach a
certain final state depend only on the distribution of $r_i$ values in
the population and on the period $S$. 
Careful consideration of eqs.~(\ref{tau_integral}) and
(\ref{tau_estimate}) which involve integration over every possible
interval of $r(s)$ leads to the conclusion that a reordering of the
function $r(s)$ would lead to the same mean consensus time. For example,
 time reversed versions of any of the
$r(s)$ functions would yield the same results.

The mean time to consensus for different flip-rate periods $S$ is
presented in fig.~\ref{fixtime_vs_S} for the three $r(s)$ functions
tried. All move from a time similar to 
that found for homogeneous $r_i$ (the original Voter Model)
for small $S$ to a value close to
that found for static $r_i$ with the same distribution (heterogeneous
Voter Model) for large $S$. 
See eqs.~(\ref{T_hetero_static}) and (\ref{T_homo_static}) in
Section \ref{analysis}.
The transition occurs over a similar range of
$S$ for each model, though the shape differs a little.

\begin{figure}[tb]
\centering\includegraphics[width=0.70\textwidth]{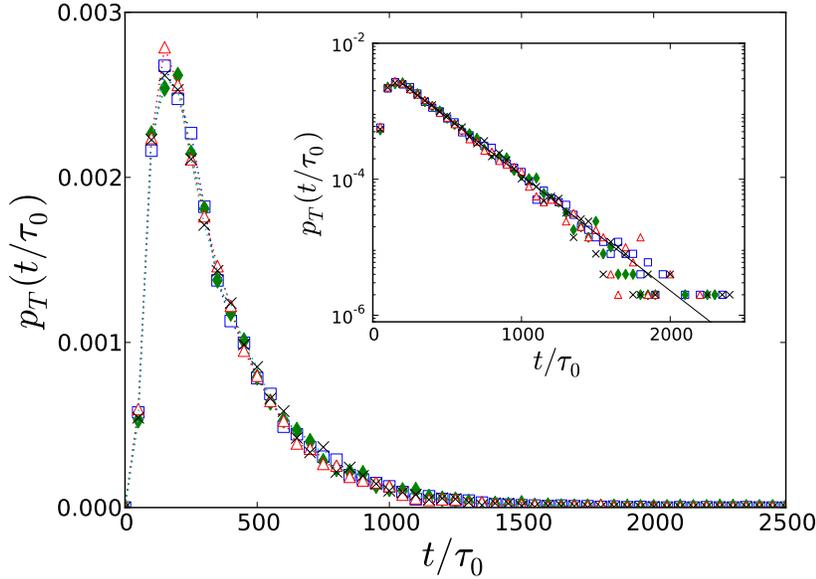}
\caption{Distribution $p_T(t/\tau_0)$ of consensus times rescaled by $\tau_0$
  for $r_{\mbox{qdr}}(s)$ and $S = 0.5$ ($\square$), $S=32$
  ($\diamond$) and $S=2048$ ($\triangle$), and for $r_{\mbox{lin}}(s)$
  with $S=128$ ($\times$). Frequencies were binned in rescaled time
  intervals of $50$. Under this time rescaling, all distributions
  collapse onto the same curve.
Inset: the same data plotted with logarithmic vertical axis, compared
with a decaying exponential function (line).
}\label{T_distn}
\end{figure}

The distribution of consensus times rises rapidly to a peak value at
small times, followed by a long tail very closely approximated by an
exponential decay. In eq.~(\ref{T_interp}) we see that the mean
consensus times for different $r$ functions or values of $S$ differ
only by the factor $\tau_0$. In fact, the whole distribution of 
consensus times has the same shape, so that if we rescale by $\tau_0$,
the distributions collapse onto the same curve, as shown in
fig.~\ref{T_distn}.
In the inset the data are plotted against a logarithmic scale, showing
the exponential tail clearly. Interestingly the decay $p_T\propto
e^{-\lambda t}$ generally does not have $\lambda = 1/T$ as might be
naively expected.

\begin{figure}[tb]
\centering\includegraphics[width=0.49\textwidth]{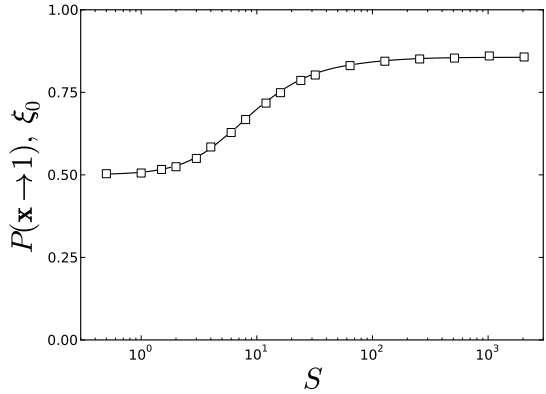}
\caption{Probability to reach consensus at the state ${\bf 1}$ as a
  function of $S$ for inhomogeneous initial conditions. Open squares
  are simulation results for $r_{\mbox{sin}}(s)$. Half the agents have
  $x_i(0) = 1$ and initial flip rate $r_{\mbox{sin}}(s_i)$ with $s_i \in
  [1/2,1)$. Remaining agents have $x_i(0) = 0$ and $s_i \in [0,1/2)$.
Data points are from $2\times10^4$ runs with $N = 500$.
Solid curve is $\tilde{\xi}_0$ calculated using
eqs. (\ref{tau_integral}) - (\ref{xi_tilde}).
}\label{fixprob_vs_S}
\end{figure}

\begin{figure}[tb]
\centering
\includegraphics[width=0.49\textwidth]{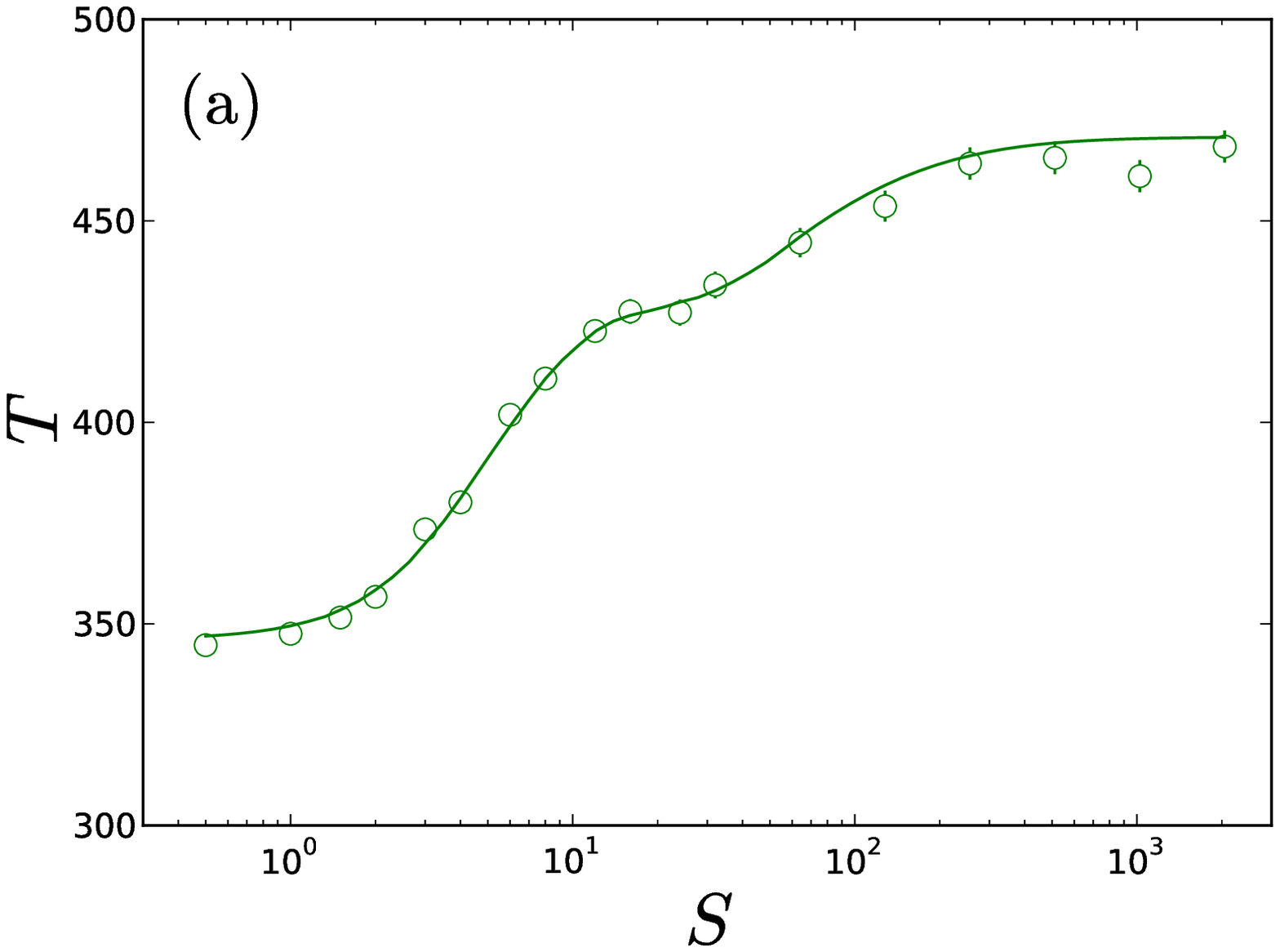}
\includegraphics[width=0.49\textwidth]{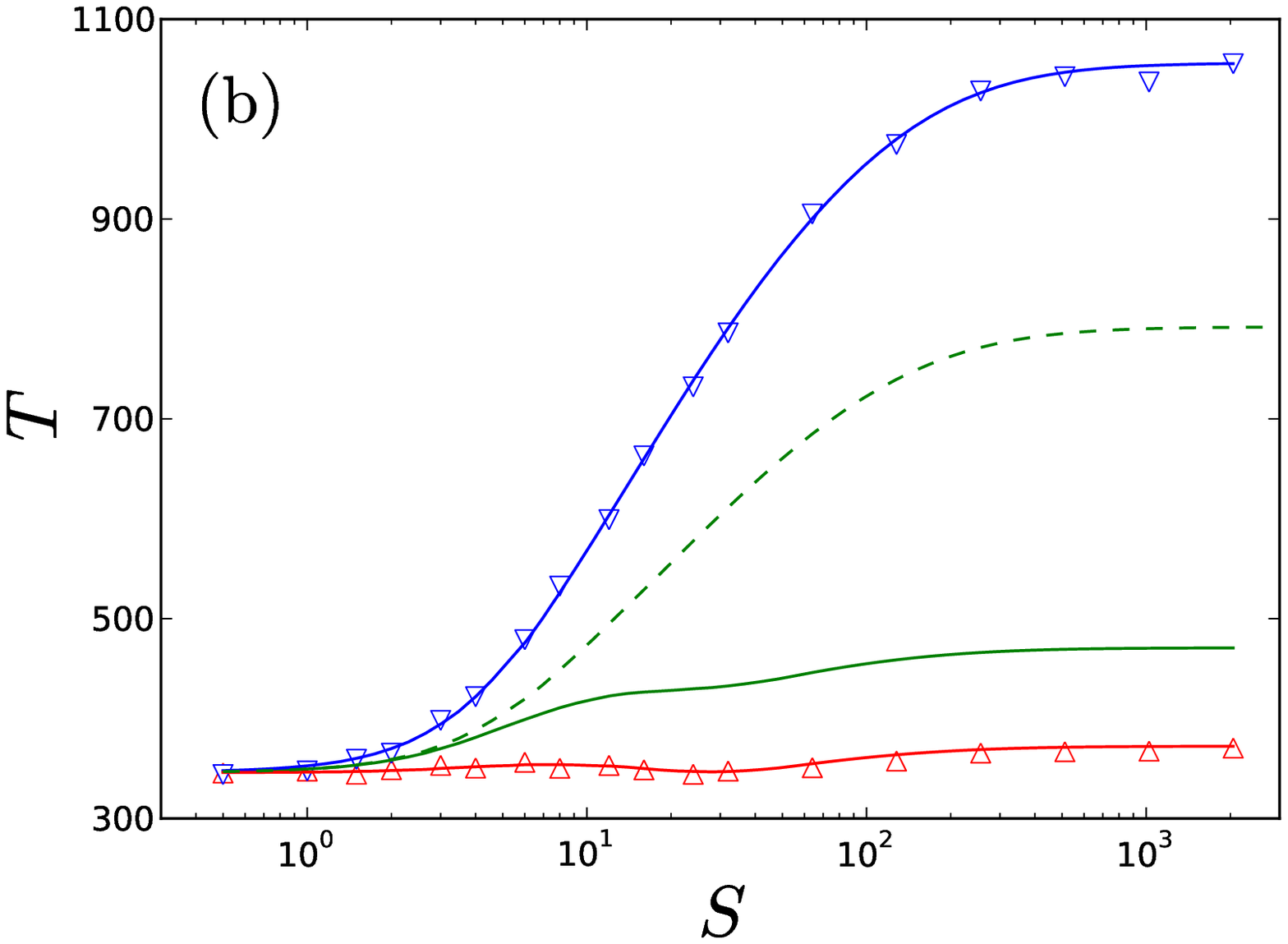}
\caption{(a) Mean consensus time as a function of $S$ for inhomogeneous
  initial conditions. Open circles
  are simulation results for $r_{\mbox{sin}}(s)$.
Solid curve is calculated consensus time using eq. (\ref{T_inhomog}).
(b) Mean time to reach each of the two possible consensus
states. ($\triangle$) mean time to reach final state ${\bf 1}$.
($\bigtriangledown$) mean time to reach final state ${\bf
  0}$. Analytical predictions of eqs.~(\ref{T0}) and (\ref{T1}) are
shown as solid lines (blue and red respectively, online).
For comparison the analytic predictions for the overall mean time to
consensus (solid, green online) from  eq. (\ref{T_inhomog})---the same
as shown in panel (a)---and for the
homogeneous case (dashed) from eq.~(\ref{T_interp}) are also shown.
Simulation conditions are the same as for fig.~\ref{fixprob_vs_S}.
}\label{fixtime_vs_S_inhomog}
\end{figure}

We also carried out simulations with inhomogeneous initial
conditions. 
Because we have so far dealt only with a simple fully-connected (or
well mixed) population, the inhomogeneity is in the correlation
between initial $r_i$ values and initial opinions.
Shown in figs.~\ref{fixprob_vs_S} and
\ref{fixtime_vs_S_inhomog} are results for $r_{\mbox{sin}}(s)$. 
The
agents were divided into two equal groups. Agents in the first group
had $x_i(0)$ set to $0$, and $s_i(0)$ chosen uniformly in
$[0,1/2)$. The agents in the second group had $x_i(0)$ set to $1$, and
  $s_i(0)$ chosen uniformly in$[1/2,1)$. In this way $x_0 = 1/2$ but
    the agents with initial opinion $1$ all had initial values of
    $r_i$ below $r_0$, while those with initial opinion $0$ all had
    $r_i(0) > r_0$. This means the probability to reach final state
    ${\bf 1}$ is
    greater than $x_0$. As can be seen in fig.~\ref{fixprob_vs_S}, the
    effect is largest for large $S$. 

As $S$ decreases, the timescale of
    change of $r_i$ eventually becomes shorter than the timescale of
    consensus, and the effect of the initial inhomogeneity is lost.
The time to reach consensus also changes with $S$, as for the
homogeneous case, but now in a more complicated way, 
 fig.~\ref{fixtime_vs_S_inhomog} (a). The inhomogeneity reduces
the overall time to
reach consensus from that found in the heterogeneous case, as can be
seen by comparing
the dotted and central solid curves in fig.~\ref{fixtime_vs_S_inhomog}
(b)
, due to the inertia of the slowly changing agents, who now initially share a
common opinion, combined with the speed with which the agents holding
the weaker opinion may be changed.

It is also interesting to compare the mean time taken to reach each of
the final states, ${\bf 0}$ and ${\bf 1}$.
The mean time to reach ${\bf 1}$ is almost as short as (but not
shorter than) that for
homogeneous $r$. The agents originally holding
opinion $1$ have a lot of weight, so the majority of agents in the
mixed QSS will have opinion $1$ and quickly kill off opinion $0$.
In the minority of cases, (see fig.~\ref{fixprob_vs_S}) the final
state is ${\bf 0}$. In this case, the time taken to reach consensus is
very long, significantly longer then the time taken for heterogeneous
$r$ with homogeneous initial conditions, as can be seen in 
fig.~\ref{fixtime_vs_S_inhomog} (b).
This has ramifications for language change, as age related variation
in flip-rates delays consensus in two ways. The mean time to reach
consensus, which corresponds to a language variant becoming
established as the convention in a population, is always longer for
heterogeneous flip-rates than for a perfectly homogeneous population,
so any changes of learning with age will delay the establishment of a
convention. The effect is exacerbated by the fact that new variants
tend to originate in the youngest members of the population
\cite{Labov01}, corresponding to opinion $0$ in the present model,
meaning the time taken for a new variant to overtake the population is
even longer.
For example, in \cite{Baxter09} the mean time to reach consensus
    among British and Irish immigrants to New Zealand
    in a feasible neutral model was found to be much longer than the
    observed time. As just described, generational effects only make
    this situation worse, suggesting that some kind of selection
    effect (preference for one variant over another) must have been at
    work in this situation.

\begin{figure}[tb]
\centering
\includegraphics[width=0.49\textwidth]{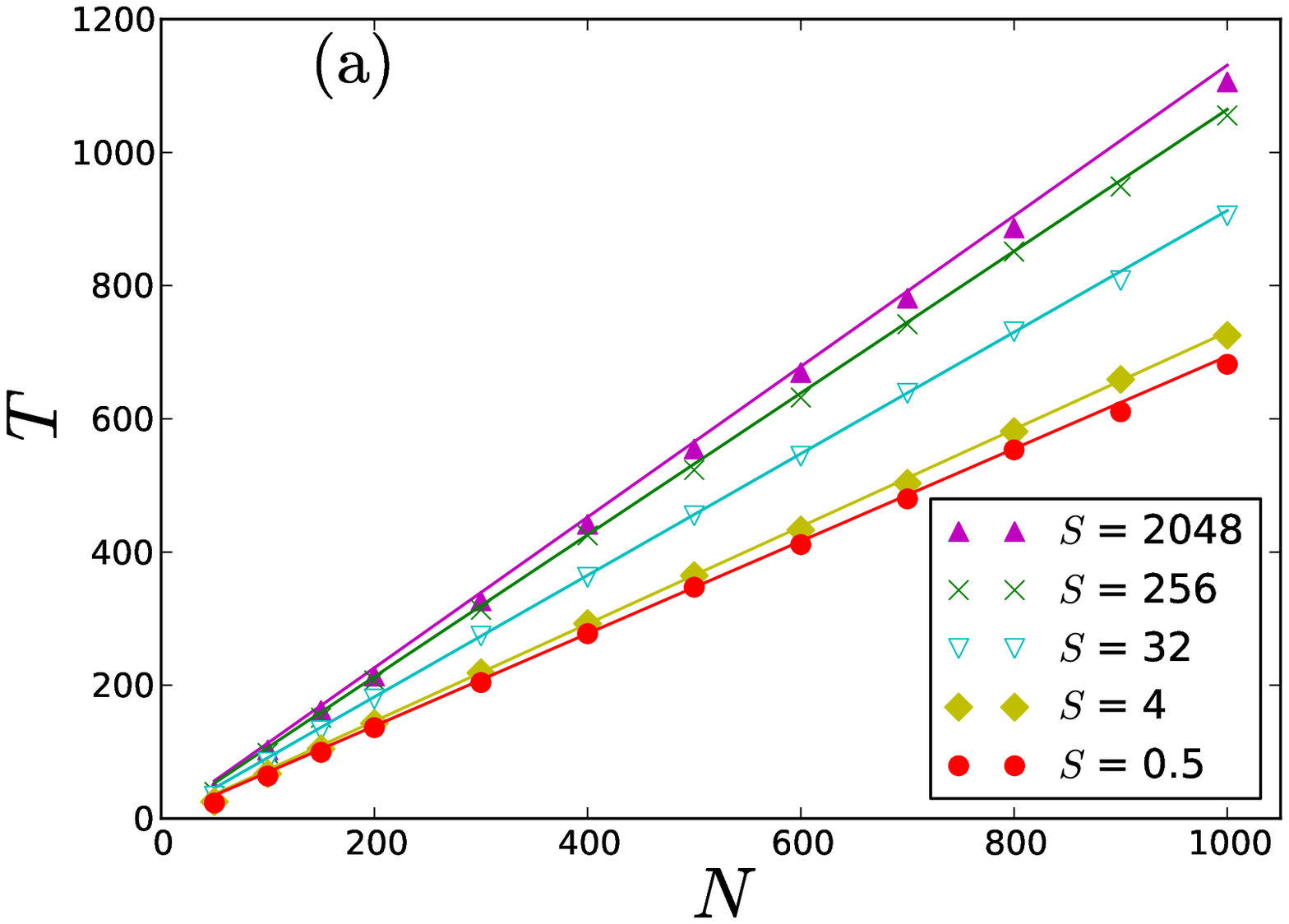}
\includegraphics[width=0.49\textwidth]{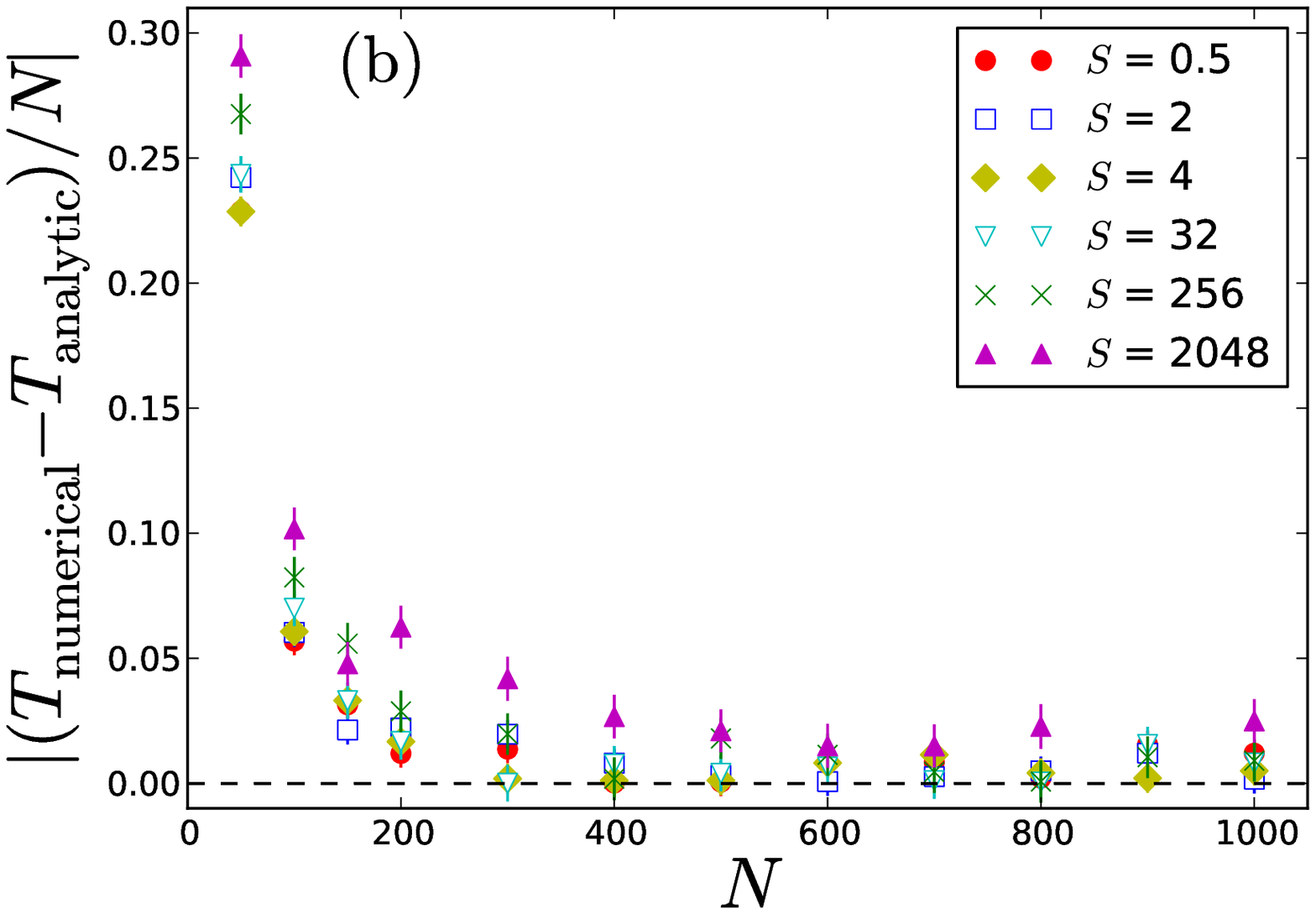}
\caption{(a) Mean consensus time versus population size for different
  values of $S$, using $r_{\mbox{lin}}(s)$. Symbols show numerical
  results, lines show predictions of eq.~(\ref{T_interp}) which grow
  linearly with $N$.
(b) The difference between numerical and analytic values for
  $T/N$ as a function of $N$ at several values of $S$. For small $S$,
  the difference is large for small $N$ but converges rapidly to
  $0$. For larger $S$ the difference also falls rapidly, but converges
  to a small yet non zero value.
}\label{error_vs_N}\label{linear_scaling}
\end{figure}

To establish the effect of system size, we
repeated numerical simulations for a range of values of $N$, from $50$
up to $1000$. 
In fig.~\ref{linear_scaling} (a) we plot $T$ as a function of $N$ for
several values of $S$ across a broad range. We see that the mean
consensus times grow linearly with $N$, confirming that the
calculated scaling $T\propto N$ is correct. Numerical results are in
excellent agreement with analytic predictions for larger $N$, given by
 by eq.~(\ref{T_interp}).
In fig.~\ref{error_vs_N}~(b) we plot the absolute difference between
the numerical and analytic results as a function of $N$, for several
$S$ values.
We see that the results for
small $N$ do not agree at all well with analytic predictions, which are
based on a large $N$ approximation. 
For
larger $N$, however, the numerical results quickly converge to the analytic
curve, coinciding for $N \geq 400$. 
For small $S$, the difference is consistent with zero for
$N$ from approximately $N=400$. For larger $S$ values, the numerical
results never exactly converge to the prediction, but the difference
achieves its minimum value from $N$ around $400$ and remains the same
as $N$ increases (aside from statistical fluctuations). All of the results
presented above are for $N=500$. Above this
value we don't expect to see any improvement in the results.

\begin{figure}[tb]
\centering\includegraphics[width=0.70\textwidth]{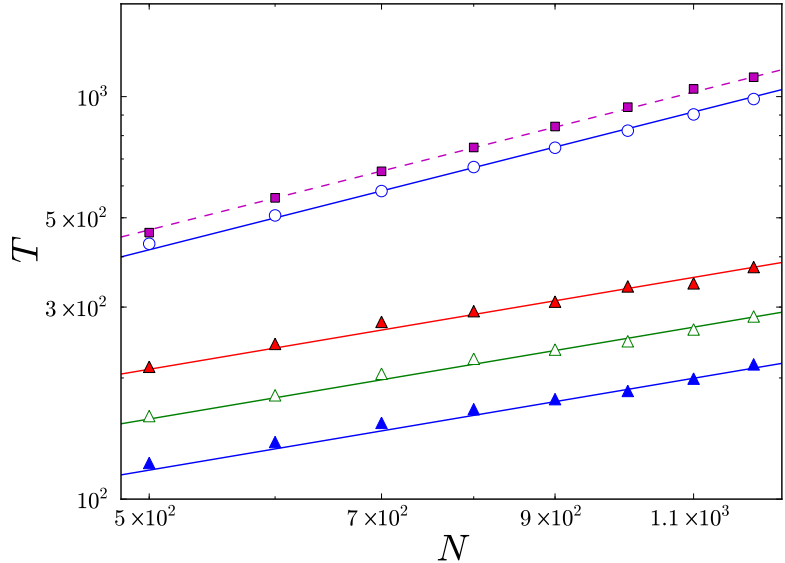}
\caption{Mean consensus time $T$ versus population size $N$ on
  different networks, plotted on logarithmic axes. From top to
  bottom: Erdos-Renyi graph with mean degree $7$ ($\square$),
  scalefree graph with exponent $\gamma=3.5$ ($\circ$) and scalefree
  graph with $\gamma=2.5$ for $S = 2048, 32$ and $0.5$ (triangles).
Solid lines are analytic predictions based on eq.~(\ref{T_interp_netw})
which predicts $T \propto N$ for Erd\H{o}s-R\'{e}nyi and scalefree with
$\gamma=3.5$ and $T \propto N^{2/3}$ for $\gamma = 2.5$.
All data are for $r_{\mbox{lin}}(s)$ with $x_0 = 0.5$ and homogeneous
initial conditions.
  }\label{networks_scaling}
\end{figure}

Finally, we extended the simulations to a population of voters on a
network. 
We carried
out simulations of the model for Erd\H{o}s-R\'{e}nyi networks and
scalefree networks, whose degree distributions follow a decaying powerlaw
of the form $P(q) \propto q^{-\gamma}$ for large degree $q$.
The qualitative behaviour is exactly the same, with mean
consensus time rising with a logistic-shaped curve from a minimum at
small $S$ to a maximum at large $S$. Equation ~(\ref{T_interp_netw})
predicts that consensus times should depend on the degree distribution
of an uncorrelated network by a factor $\langle q\rangle^2 /\langle
q^2\rangle$. In general, this factor does not depend on network size,  meaning consensus times will grow linearly
with $N$. If $\langle q^2\rangle$ diverges with $N$, a different scaling
of mean consensus time with $N$ will emerge. 
For scalefree networks with 
$\gamma <3$, $\langle q^2\rangle \propto
N^{(3-\gamma)/(\gamma-1)}$\footnote{We did not restrict multiple edges
  in this simulation.}
meaning we expect to find $T \propto N^{(2\gamma-4)/(\gamma-1)}$. 
In fig.~\ref{networks_scaling} we plot mean
consensus time for various $N$ for an  Erd\H{o}s-R\'{e}nyi network
with mean
degree $7$, and scalefree networks with $\gamma = 3.5$ and $\gamma =
2.5$. As expected, $T$ grows linearly with $N$ for the Erd\H{o}s-R\'{e}nyi
network and scalefree network with $\gamma = 3.5$. For $\gamma = 2.5$
the mean consensus time grows sublinearly with $N$, with an exponent
close to the expected value of $2/3$.

\section{Discussion\label{discussion}}

In this paper we have introduced a generalisation of the Voter Model
in which the flip-rates of agents vary with time. Interaction between
the time scale of consensus formation and the time scale of flip-rate
change leads to non-trivial dependence of mean consensus time on the
period of flip-rate change. For very rapidly changing flip-rates, the
mean time to consensus agrees with that found for the Voter Model with
fixed homogeneous flip-rates.
As the period of change of the flip-rates lengthens, so does the
consensus time, until it saturates at the time found for static
heterogeneous flip rates. 
An analytic estimate of the mean consensus time can be found
by calculating the expected interval between interactions for
each agent, then using the
usual method of assuming a quickly reached quasi-stationary state
followed by a slow escape to consensus. The
 mean consensus time is calculated for this second stage through a
 Fokker-Planck equation for a single
conserved centre-of-mass variable.
The results obtained by this method are in excellent agreement with
numerical simulation.
The overall mean time to consensus, the mean time to reach a
particular final state and the rescaling of the distribution of
consensus times are all correctly predicted. We also found that the 
 complex network
structures such as scale-free networks affect the scaling of the mean
consensus time with population size
in the same way that they do for static flip-rates.

For simplicity here we used periodic
flip-rates, but the method used is readily applicable to more complex
variations in flip-rates.
For example, if each agent's
flip-rate varies with a different period, or indeed if each followed a
different function entirely.
More generally, time dependent interactions in any copying process
may be modelled and analysed in a similar way, to consider for example
seasonal variation in invasion rates in ecological models, or
time-variation in
the strength of synaptic interactions in neuronal models.
This is particularly relevant to 
language change. Speakers learn more quickly when they are young and
more slowly or almost not at all once they reach adulthood. The effect
of such ageing can be modelled by exactly the kind of time-varying
interactions described here. The results presented here for this very
simple model suggest that in a more realistic language change model,
heterogeneity in learning rates will increase the mean time taken
to reach consensus, and this will be further exacerbated as a new language
variant is more likely to appear in the faster learning (i.e. the
younger) members of the population. The inertia of the (older) slowly
adapting speakers will contribute both to enhanced survival probability
for the existing convention, and in the event that a new variant does
take over to increasing the time required for this to happen.
Extension of time variation of update rates to such more realistic
models is therefore a natural avenue
for future investigation.
Another possible extension would be to consider flip-rates that depend
on local dynamical
processes, which is relevant for example in the case of neuronal
models in which a neuron's response depends on recent activity.

\ack

This work was partially supported by the Portuguese Foundation for
Science and Technology project PTDC/FIS/71551/2006 and post-doctoral
fellowship SFRH/BPD/74040/2010. 
I thank R.~Blythe and A.~McKane for helpful comments on the
manuscript, and M.~Barroso for 
the administration of the \verb=blafis= computing facilities used for
numerical simulations. 

\medskip

\bibliography{voter}

\end{document}